\documentclass[twocolumn,showpacs,preprintnumbers,nofootinbib]{revtex4}

\usepackage{latexsym,amssymb,amsmath,bbm,graphicx,here,epsfig}
\usepackage{psfrag} %replace labels in postscript files

\allowdisplaybreaks[1] %for amsmath package

\newcommand{\STr}{\mathrm{STr}}

\newcommand{\Bs}{\boldsymbol}

\renewcommand{\bar}{\overline}

\begin{document}
%%%%%%%%%%%%%%%%%%%%%%%%%%%%%%%%%%% revtex title %%%%%%%%%%%%%%%%%%%%%%%%%%%%%%%%%%
\preprint{HD-THEP-03-34}
\title{Antiferromagnetic gap in the Hubbard model}

\author{Tobias Baier}
\author{Eike Bick}
\author{Christof Wetterich}%
\affiliation{%
  Institut  f\"ur Theoretische Physik\\
  Universit\"at Heidelberg\\
  Philosophenweg 16, D-69120 Heidelberg
}%

\begin{abstract}
We compute the temperature dependence of the antiferromagnetic order parameter and
the gap in the two dimensional Hubbard model at and close to half filling.
Our approach is based on truncations of an exact functional renormalization group
equation. The explicit use of composite bosonic degrees of freedom permits a direct
investigation of the ordered low temperature phase. We show that the Mermin--Wagner
theorem is not practically applicable for the spontaneous breaking of the continuous
spin symmetry in the antiferromagnetic state. The critical behavior is dominated by
the fluctuations of composite Goldstone bosons.
\end{abstract}

\maketitle

The low temperature phase of the Hubbard model \cite{Hub63} is highly interesting
for a variety of phenomena observed in strongly correlated electron systems. This
includes the antiferromagnetic state for zero or small doping, on which we concentrate
in this note, as well as possible superconducting states for large doping. There has been
a long debate about the effective theory for the spin wave fluctuations
\cite{ETD1}, \cite{ETD2} and the correct implementation of mean field theory
\cite{MFD1,MFD2,MFD3}, the problem being to reconcile the SO(3)-spin symmetry with the
Hartree-Fock results. More recently, functional renormalization group studies have made
progress in the understanding of instabilities in various channels and the corresponding
phase diagram \cite{Zan97,Hal00,Hon01,Gro01}, but have been unable to explore the ordered
phases for low temperature. (For other approaches see
\cite{VD,Sca95}.) We concentrate in this note
on the two dimensional Hubbard model with next neighbor interactions.
There a further problem arises for the understanding
of the antiferromagnetic state - a theorem by Mermin and Wagner \cite{Mer66} forbids the
spontaneous breaking of a continuous symmetry at nonvanishing temperature. This seems to
contradict an antiferromagnetic order since the latter breaks spontaneously the
continuous spin symmetry.

In this letter we present a combined study of Schwinger-Dyson (or Hartree-Fock) equations
\cite{SD,HF}, extended mean field methods \cite{Bai00}, \cite{Bai01} using a generalized
Hubbard-Stratonovich transformation \cite{HubStrat} and functional renormalization
based on an exact renormalization group equation \cite{Wet93a,Ber02} for the
effective average action \cite{Wet91}. We propose a partial bosonization which respects all
symmetries. For a suitable choice the mean field theory (where only the fermionic
fluctuations are taken into account)
reproduces the Hartree-Fock result. For generic choices, however, the mean field phase
diagram depends on unphysical parameters \cite{Bai00}, reflecting the ``Fierz ambiguity''
\cite{Jae02} in the formulation of  mean field theory. This problem is overcome by the
inclusion of the bosonic fluctuations. The size of the bosonic corrections depends
in such a way on the choice of partial bosonization that the phase diagram
becomes almost independent of the choice of the mean field already for the approximations
employed in the present work \cite{BBWL}.

The scale dependence of the effective action is followed by the solution of a
functional flow equation. This is formulated in the partially bosonized theory,
employing the ``rebosonization'' technique of \cite{Gie02}.
We thus combine the capability of mean field theory to
describe an order parameter with the advantages of the functional renormalization group.
The renormalization flow includes step by step the bosonic fluctuations on larger and
larger length scales and the fermionic fluctuations closer and closer to the Fermi surface
as the ``infrared cutoff'' $k$ is lowered. This formulation of the functional
renormalization group can be used in the low temperature phase as well, in contrast
to the purely fermionic formulation \cite{Zan97,Hal00,Hon01,Gro01}.

A coherent picture for the antiferromagnetic phase at low temperature and small doping
emerges. We find a simple effective action for the spin waves (cf. eq. (\ref{eq:trunk_bos_I})
and a more detailed discussion in \cite{BBWL}) which respects all symmetries. For
temperatures near and below the effective critical temperature
$T_c$ and for small doping (small chemical
potential $\mu$) it can be described by an $O(3)$-symmetric effective potential
$U(\vec{a})$ for the antiferromagnetic order parameter $\vec{a}$ and an appropriate
kinetic term which reads for small momenta $(Z_at^2/2)\partial_i\vec{a}\partial_i\vec{a}$.
(Here $Z_a$ is a wave function renormalization, $t$ is the next neighbor coupling of the
Hubbard model and the derivatives $\partial_i$ are in lattice units.) For $T<T_c$
the minimum of $U$ occurs for $\alpha_0=\vec{a}^2_0/2\neq 0$. The characteristic
scale for the antiferromagnetic order is set by the renormalized order parameter
$\kappa_a=Z_at^2\alpha_0/T$. In particular, this determines the gap for the electrons,
$\Delta_a=\sqrt{2T\kappa_a/Z_a}(\bar{h}_a/t)$, with Yukawa coupling $\bar{h}_a/t$ close
to one for our numerical example \cite{BBWL}. Also the renormalized mass (or inverse
correlation length) for the spin fluctuations in the ``radial direction'' (parallel to the
order parameter $\vec{a}_0$) is determined by $\kappa_a,~\tilde{m}_a\sim \sqrt{\kappa_a}$,
whereas the spin fluctuations in the ``Goldstone direction'' perpendicular to $\vec{a}_0$
are massless (infinite correlation length) for all $T<T_c$. Our result for the order
parameter $\kappa_a$ is plotted in fig. 1.

%%%%%%%%%%%%%%%%%%%%%%%%%%%%%%%%%%begin{figure}%%%%%%%%%%%%%%%%%%%%%%%%%%%%%%%%%%
\begin{figure}[htb]
\centering
\psfrag{T}{$T/t$}
\psfrag{kappa}{$\kappa$}
\includegraphics[scale=0.65]{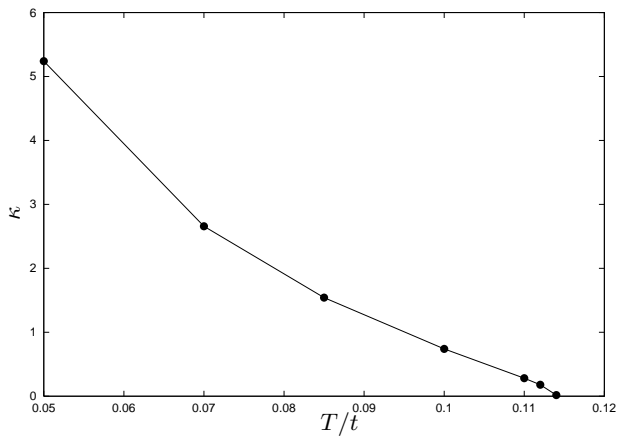}

\caption{Renormalized expectation value $\kappa_a$ of the
antiferromagnetic condensate in the low temperature phase
for $\mu=0, U/t=3$.}
\label{fig:ssb}
\end{figure}
%%%%%%%%%%%%%%%%%%%%%%%%%%%%%%%%%%end{figure}%%%%%%%%%%%%%%%%%%%%%%%%%%%%%%%%%%%%

Actually, as a particularity of the two dimensional model, the size of the gap
$\Delta_a$ (and the value of $\kappa_a$) depends on the infrared scale $k$ which
may be associated to the inverse of the size of the experimental probe.
(We take $k^{-1}\approx 1cm$.) This is due to the fluctuations of
the massless Goldstone bosons which reduce $\Delta_a$ as $k$ is lowered. As we can
see in fig. 2 this decrease is only logarithmic. For any fixed $T\neq0$ and $\mu$ one
always encounters a scale $k>0$ where $\kappa_a$ vanishes. Therefore, the order parameter
vanishes for $k\rightarrow 0$, in agreement with the Mermin-Wagner theorem.
Nevertheless, for low enough $T$ the effective order parameter remains nonzero for any
realistic macroscopic length $k^{-1}$. In this sense the Mermin-Wagner theorem fails to be
practically applicable to the low temperature behavior of the two dimensional Hubbard model.
For fixed macroscopic $k^{-1}$ and $T<T_c$ the latter
is well described by an ordered antiferromagnetic phase.

\begin{figure}[tb!]
\centering
\psfrag{-t}{\hspace{-1.5cm}$-\ln k/t$}
\includegraphics[scale=.55]{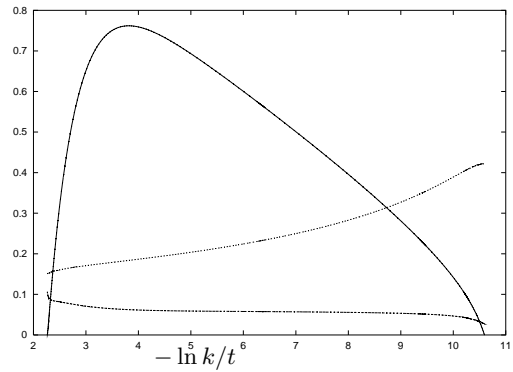}
\caption{Flow of the renormalized minimum of the potential $\kappa_a$ (solid),
the quartic bosonic coupling $10^{-2}\lambda_a$ (long dashes) and the wave function
renormalization $10^{-1}Z_a$ (short dashes). We have chosen $U/t=3$ and $T/t=0.15;~\mu=0$.}
\label{fig:ayuk_ssb}
\end{figure}

The effective critical temperature $T_c$ depends mildly on the size of the probe $k^{-1}$ -
a change of $k$ by a factor ten shifts $T_c$ by $10\%$. Nevertheless, the two dimensional
model exhibits effective critical behavior near $T_c$, with a macroscopic correlation
length. For $k \ll T$ the fermion fluctuations effectively
decouple and the flow is described by
a classical $O(3)$ linear scalar model \cite{Ger01} (right part of fig. \ref{fig:ayuk_ssb}).
For the correlation length $\xi$ we find for $T>T_c$
\begin{equation}\label{AA}
\xi(T)=\frac{c(T)}{t}\exp\left\{20.7\beta(T)\frac{T_c}{T}\right\}
\end{equation}
with $T_c=0.114t~,~\beta(T_c)=1$ and $c(T),\beta(T)$ smoothly
varying functions of $T$ \cite{BBWL}
without particular features for $T\rightarrow 0$. This closely
resembles the result of \cite{CHN} which would  correspond to constant $\beta$. Near
$T_c$ the correlation length reaches the macroscopic size of the probe.
Eq. (\ref{AA}) is no longer applicable for $T<T_c$.

The antiferromagnetic order occurs in the
low temperature phase for a whole range of small doping. The critical temperature
decreases mildly with the chemical potential
$\mu$ \cite{BBWL}. We emphasize that the quantitative
impact of the Goldstone-boson fluctuations on the value of $T_c$ is very substantial.
One may define a ``pseudocritical temperature'' $T_{pc}$ where the mass term for the
antiferromagnetic spin wave reaches zero for some $k$, corresponding in the fermionic
language to the four fermion coupling in the associated momentum channel growing
``infinitely'' large. We find $T_{pc}/t=0.18(0.21)$ for $\mu=0,T_{pc}/t=0.145(0.18)$
for $\mu/t=0.15$ (with brackets for mean field theory) - to be confronted with
$T_c=0.115$ for $\mu=0$.

%%%%%%%%%%%%%%%%%%%%%%%%%%%%%%%%%%begin{figure}%%%%%%%%%%%%%%%%%%%%%%%%%%%%%%%%%%
%%%%%%%%%%%%%%%%%%%%%%%%%%%%%%%%%%end{figure}%%%%%%%%%%%%%%%%%%%%%%%%%%%%%%%%%%%%
\vspace{0.3cm}

We next describe our formalism and approximations - for more details see \cite{BBWL}.
We work within a functional integral formalism where the coordinates $X$ denote the
lattice sites as well as a ``Euclidean time'' $\tau$. We concentrate in this note on
fermion bilinears corresponding to spin density
waves $\tilde{\vec{m}}$ and charge density waves $\tilde{\rho}$.
\begin{equation}
\tilde{\vec m}(X)=\hat\psi^\dagger(X)\vec\sigma\hat\psi(X)~,~
\tilde\rho(X)=\hat\psi^\dagger(X)\hat\psi(X).
\end{equation}
With these bilinears the four fermion coupling of the Hubbard model can be written as
\begin{equation}
  (\hat\psi^\dagger(X)\hat\psi(X))^2=\tilde\rho(X)^2=-\tfrac 13\tilde{\vec m}(X)^2.
\end{equation}
We define the partially bosonized partition function by
\begin{equation}
  \label{eq:partition_bos}
  Z[\eta,\eta^*,J_\rho,\vec J_m] = \int {\cal D}(\psi^*,\psi,\hat\rho,\hat{\vec m})\exp \big( - (S+S_\eta+S_J) \big)
\end{equation}
with bosonic fields $\hat{\vec{m}},\hat{\rho}$, sources $\eta,J$ and classical action
\begin{equation}
  \label{eq:action_bos}
  \begin{gathered}
    S=S_{F,\text{kin}}+\tfrac 12 U_\rho\hat\rho^2+\tfrac 12 U_m\hat{\vec m}^2-U_\rho\hat\rho\tilde\rho-U_m\hat{\vec m}\tilde{\vec m},\\
    S_\eta = -\eta^\dagger\psi-\eta^T\psi^*, \quad S_J = -J_\rho\hat\rho-\vec J_m\hat{\vec m}.
  \end{gathered}
\end{equation}
In particular, the fermion kinetic term in momentum space takes the form $\big(\omega_F=(2n+1)\pi T\big)$
\begin{eqnarray}
S_{F,\text{kin}}&=&\sum_Q\hat\psi^\dagger(Q)P_F(Q)\hat\psi(Q),\\
P_F(Q)&=&i\omega_F+\epsilon-\mu, \quad \epsilon(\Bs q) = -2t(\cos q_x+\cos q_y)\nonumber
\end{eqnarray}
such that the Fermi surface is located at $-2t(\cos q_1+\cos q_2)=\mu$. Here the
Fourier transforms are defined as $\hat\psi(X)=\sum_Q\hat\psi(Q)e^{iQX}$
and we use the short hand notation
$Q=(\omega_F,\Bs{q}),\quad X=(\tau,\Bs{x}),\quad QX=\omega_F\tau+\Bs{x}\Bs{q}$.

The Gaussian functional integral over the spin density boson $\hat{\vec{m}}(x)$ and
charge density boson $\hat{\rho}(x)$ can be performed explicitely. This demonstrates that
$W=\ln Z$ is equivalent to the free energy of the Hubbard model with next neighbor coupling
and repulsive interaction $U>0$ (up to terms $\sim J^2$ which are explicitely calculable).
The equivalence holds for all $U_\rho$ and $U_m$ subject to the constraint
\begin{equation}
  \label{eq:coupcond}
  U=-U_\rho+3U_m.
\end{equation}
The remaining freedom in the choice of $U_m$ is related to the Fierz ambiguity. The
Hartree-Fock or lowest order Schwinger Dyson
approximation corresponds to a computation of the fermionic fluctuation determinant
(neglecting the fluctuations of $\hat{\rho},~\hat{\vec{m}}$) for the choice $U_\rho=U_m
=U/2$. (See \cite{Bai00} for a
generalized setting).

The renormalization flow is implemented by the introduction of an infrared cutoff $k$ through
a piece in the action quadratic in the fields
$\hat{\chi}=(\psi,\hat{\rho},\hat{\vec{m}})=(\psi,\hat{b})$. In
presence of this cutoff the generating functional for the connected Green functions depends on the scale $k$
\begin{eqnarray}\label{eq:defWk}
W_k[J] &=& \ln\int{\cal D}\hat\chi \exp\big(-(S[\hat\chi] +
\Delta S_k[\hat\chi]) + J\hat\chi)\big),\\
\Delta S_k&=&\sum_Q\big\{\hat{\psi}^\dagger (Q)R^\psi_k(Q)\psi(Q)
+\frac{1}{2}\hat{b}(-Q)R^b_k(Q)\hat{b}(Q)\big\}.\nonumber
\end{eqnarray}
For the fermionic cutoff we choose
\begin{equation}
  R_k^\psi(Q) = i\omega_F(\tfrac{T_k}{T}-1) = 2\pi i(n_F + \tfrac 12)(T_k-T),
\end{equation}
which has the effect of replacing the temperature $T$ by some function $T_k$ in the
fermionic propagator. We specify this function by $T_k^4 = T^4 + k^4$,
such that for large $k$ the flow mimics a change in temperature whereas for $k\ll T$ the modification of the fermion propagator becomes ineffective.
For the bosonic cutoff we take

\begin{equation}
  \label{eq:bos_regulator}
  R_k^b(Q) = Z(k^2-\hat Q^2) \Theta(k^2-\hat Q^2),
\end{equation}
where $\hat Q^2=Q^2$ in an appropriate interval, see
eq. \eqref{eq:trunk_bosprop}.

We define the {\em effective average action} as $(\chi=\partial W/\partial J)$
\begin{equation}
  \label{eq:def_AEA}
  \Gamma_k[\chi] = J\chi-W_k[J]-\Delta S_k[\chi].
\end{equation}
The dependence on the scale $k$ obeys an exact flow equation \cite{Wet93a}
\begin{equation}\label{eq:flowAEA}
\partial_k \Gamma_k[\chi]
= \tfrac 12 \STr\big\{ \partial_k R_k [\Gamma^{(2)}_k+R_k]^{-1} \big\},
\end{equation}
which has a physically intuitive ``one loop form'' \cite{Wet91}. Here
the ``supertrace'' runs over field type, momentum and internal indices
and has an additional minus sign for fermionic entries. We see that the IR cutoff $R_k$
adds to the exact inverse propagator matrix $\Gamma^{(2)}_k$ (second functional
derivative) and therefore regulates possible zero modes, removing potential infrared
problems for any $k>0$. The effective average action interpolates between the classical
action $S$ (``initial condition'' for $k\rightarrow \infty$) and the full ``quantum
effective action'' for $k\rightarrow 0$.

As we are mainly interested in antiferromagnetic behavior we define the boson
$\big(\Pi=(0,\pi,\pi)\big)$
\begin{equation}
\vec a(Q) = \vec m(Q+\Pi)~,~\vec{a}(X)=(-1)^P\psi^\dagger(X)\vec{\sigma}\psi(X),
\end{equation}
with $P=1$ for odd lattice sites and $P=0$ else. The antiferromagnetic order corresponds to
$\vec{a}(X)$ independent of $X$.

For an approximative solution of the functional differential equation (\ref{eq:flowAEA})
we employ a truncation of $\Gamma_k$ consisting of a fermionic kinetic term, a Yukawa
interaction between the fermions and bosons and a bosonic effective action for the
$\vec{a}$-boson. We describe here the choice $U_\rho=0~,~U_m=U/3$ whereas an
extended truncation including the fluctuations of the $\rho$-bosons can be found in
\cite{BBWL}. For the fermionic kinetic term we adopt the classical part unchanged
$\Gamma_{\psi,k}=S_{F,kin}(5)$.
The Yukawa coupling $\bar h_{a,k}$ is taken to be scale dependent
\begin{eqnarray}
\label{eq:trunk_yuk_I}
\Gamma_{Y,k}[\psi,\psi^*,\vec a] =-\bar h_{a,k}\sum_{KQQ'}
\vec a(K)\psi^*(Q)\vec\sigma\psi(Q')\\
\times\delta(K-Q+Q'+\Pi)\nonumber
\end{eqnarray}
and the purely bosonic part is described by a kinetic term and an effective potential
\begin{equation}
  \label{eq:trunk_bos_I}
  \Gamma_{a,k}[\vec a] = \tfrac 12\sum_Q\vec a(-Q) P_a(Q)\vec a(Q) + \sum_X U_k
  \big(\alpha(X)\big).
\end{equation}
Due to $SO(3)$ spin symmetry the $k$-dependent effective
potential $U_k$ only depends on the rotation invariant combination
$\alpha(X) = \frac 12 \vec a(X)\vec a(X)$. The boson kinetic term $(\omega_B=2\pi mT)$
\begin{equation}
  \label{eq:trunk_bosprop}
  P_a(Q) = Z_{a,k} \hat Q^2 = Z_{a,k} (\omega_B^2 + t^2 [\Bs q]^2),
\end{equation}
involves a scale dependent wave function renormalization $Z_{a,k}$ (with the dimension of
$\text{mass}^{-1}$) and the function $[\Bs q]^2$ is defined as $[\Bs q]^2 = q_x^2 + q_y^2$
for $q_i\in [-\pi,\pi]$ and continued periodically otherwise.

The flow equation for the effective potential $U_k(\alpha)$ obtains by evaluating eq.
\eqref{eq:flowAEA} for a homogeneous antiferromagnetic order parameter,
$\vec a(Q) = \vec a \delta(Q)$, $\alpha=\frac 12 \vec a^2$. The contribution of
the fermionic fluctuations can be found by replacing in the mean field theory result
$T\to T_k$ in the fermionic propagator and applying the formal derivative
$\tilde{\partial}_k=(\partial T_k/\partial k)\partial/\partial T_k$.
The bosonic contribution is the same as for the
$O(3)$ linear $\sigma$--model \cite{Wet91,Ber02}:
\begin{eqnarray}\label{eq:EffPotA}
\partial_k U  (\alpha) &=&
- 2T\int_{-\pi}^{\pi}\frac{d^2q}{(2\pi)^2} \tilde\partial_k\ln\cosh y(\alpha)\\
&&+\frac{1}{2}\sum_{Q,i}\tilde\partial_k \ln[P_a(Q)+\hat M_i^2(\alpha)+R^a_{k}(Q)]\nonumber
\end{eqnarray}
Here the squared fermion mass (gap)
$\Delta^2_a=2\bar h_a^2\alpha$ enters the function
\begin{equation}
y(\alpha) = \frac{1}{2T_k}\sqrt{\epsilon^2(\Bs q) + 2\bar h_a^2 \alpha}.
\end{equation}
The boson mass terms $\hat M_i^2$ obtain from the second derivatives of $U_k$ and
are different for the ``radial mode'' (in the direction of $\vec a$) and the ``Goldstone
modes'' (perpendicular to $\vec a$).
In the SSB regime we recognize two massless Goldstone bosons for $\alpha$ at the minimum
of the potential, $\alpha=\alpha_0\neq 0$.

The flow of $Z_a$ is determined by the anomalous dimension $\eta_a=-k\partial_k\ln Z_a$,
\begin{eqnarray}\label{20}
&&\eta_a = h_a^2 t k(\partial_kT_k)\Big[\frac{\partial}{\partial T_k}
\frac{\partial}{\partial (l^2)}\\
&&\Big\{\frac{1}{T_k}\int_{-\pi}^{\pi} \frac{d^2q}{(2\pi)^2}
\frac{\tanh\tfrac{\epsilon(\Bs q)+\mu}{2T_k}
+\tanh\tfrac{\epsilon(\Bs q+l\hat e_1)-\mu}{2T_k}}
{(\epsilon(\Bs q)+\epsilon(\Bs q+l\hat e_1))/t}
\Big\}\Big]_{l=0}.\nonumber
\end{eqnarray}

For the running of the renormalized Yukawa coupling $h_a=(T/Z_a)^{1/2}t^{-2}\bar{h}_a$
we obtain in the symmetric regime $(\alpha_0=0)$ a direct and a ``rebosonized''
contribution

\vspace{0.1cm}
\begin{eqnarray}\label{eq:floweq_ha}
k\partial_k h_a^2 &=& h_a^2 \eta_a + \beta_{h^2_a}^{(d)} + \beta_{h^2_a}^{(rb)},\nonumber\\
\beta_{h^2_a}^{(d)}
&=& - 2 h_a^4 \frac{t^2}{k T} \tilde\partial_k \sum_Q
\big\{ p_a^{-1}(Q) p_F^{-1}(Q)p_F^{-1}(Q+\Pi) \big\} \nonumber\\
\beta_{h^2_a}^{(rb)}
&=& 2m_a^2 h_a^4 \frac{t^2}{k T} \tilde\partial_k\sum_Q\big\{p_a^{-1}(Q)p_a^{-1}(\Pi-Q)
\nonumber\\
&&p_F^{-1}(-Q)[p_F^{-1}(\Pi-Q)-p_F^{-1}(Q)]\big\}
\end{eqnarray}
with $m^2_a=\partial U_k/\partial\alpha_{|\alpha=0}~,~
p_F(Q) = (i\hat{\omega}_F-\mu+\epsilon({\bf q}))/t$,~ $\hat{\omega}_F=(2n+1)\pi T_k$,~
$p_a(Q) = Q_k^2/k^2 + m_a^2~,~Q^2_k=\hat{Q}^2\Theta(Q^2-k^2)+k^2\Theta(k^2-\hat{Q}^2)$.
In the SSB-regime $(\alpha_0\neq 0)$ the change of the Yukawa
coupling is negligible as we have checked numerically.
Our results obtain from a numerical solution of the flow equations (\ref{eq:EffPotA}),
(\ref{20}), (\ref{eq:floweq_ha}) with a quartic approximation to $U_k(\vec{a})$
\cite{BBWL}. The initial conditions
for very large $k$ are given by $\bar{h}_a=U_m=U/3,Z_a=0,U_k(\alpha)=U_m\alpha$.

We believe that
our truncation catches the most important features in the region of small doping.
Our method gives a unified description which covers at once several interesting
limiting cases: perturbation theory and the perturbative renormalization flow of the
four-fermion coupling (small $U$), Hartree-Fock $(Z_a=0)$ and the classical spin wave
description $(k\ll T)$. The universal long-distance features (e.g. eq. (\ref{AA}))
are quite robust and should apply to a large class of models, including realistic
materials: for decreasing $T$ the correlation length increases so rapidly that it reaches
the macroscopic size of the probe $k^{-1}$ at $T_c$. For $T<T_c$ the formal symmetry
restoration would require the randomization of antiferromagnetic domains with size much
larger than $k^{-1}$. This effect is absent in a real experiment and macroscopic order can
therefore be observed.

The analysis can be extended to include other bosonic channels,
including the superconducting channel relevant for large doping \cite{Bai00}, \cite{BBWL}.
This would also lead to a more complete treatment of the momentum dependence of the
effective four fermion interaction.

%%%%%%%%%%%%%%%%%%%%%%%%%%%%%%%%%%%% bibliography %%%%%%%%%%%%%%%%%%%%%%%%%%%%%%%%%%%%

%%%%%%%%%%%%%%%%%%%%%%%%%%%%%%%%%%%%%%%%%%%%%%%%%%%%%%%%%%%%%%%%%%%%%%%%%%%%%%%%%%%%%%

\end{document}